\documentclass[12pt,titlepage]{utarticle}

\DeclareMathSymbol{\square}{\mathord}{AMSa}{"03}
\DeclareMathSymbol{\rightsquigarrow}{\mathrel}{AMSa}{"20}

\newdimen\tableauside\tableauside=1.0ex
\newdimen\tableaurule\tableaurule=0.4pt
\newdimen\tableaustep
\def\phantomhrule#1{\hbox{\vbox to0pt{\hrule height\tableaurule width#1\vss}}}
\def\phantomvrule#1{\vbox{\hbox to0pt{\vrule width\tableaurule height#1\hss}}}
\def\sqr{\vbox{%
  \phantomhrule\tableaustep
  \hbox{\phantomvrule\tableaustep\kern\tableaustep\phantomvrule\tableaustep}%
  \hbox{\vbox{\phantomhrule\tableauside}\kern-\tableaurule}}}
\def\squares#1{\hbox{\count0=#1\noindent\loop\sqr
  \advance\count0 by-1 \ifnum\count0>0\repeat}}
\def\tableau#1{\vcenter{\offinterlineskip
  \tableaustep=\tableauside\advance\tableaustep by-\tableaurule
  \kern\normallineskip\hbox
    {\kern\normallineskip\vbox
      {\gettableau#1 0 }%
     \kern\normallineskip\kern\tableaurule}%
  \kern\normallineskip\kern\tableaurule}}
\def\gettableau#1 {\ifnum#1=0\let\next=\null\else
  \squares{#1}\let\next=\gettableau\fi\next}
\def\Pf{{\rm Pf ~}}

\begin{document}

\preprint{
 HUB-EP-97/20\\
 {\tt hep-th/9703172}\\
}

\title{Seiberg duality in three dimensions}
\author{Andreas Karch
 \thanks{Work supported by DFG}
 \oneaddress{
  \\
  Humboldt-Universit\"at zu Berlin\\
  Institut f\"ur Physik\\
  D-10115 Berlin, Germany\\
  {~}\\
  \email{karch@qft1.physik.hu-berlin.de}
 }
}
\date{March 25, 1997}

\Abstract{
We analyze three dimensional gauge theories with $Sp$ gauge group.
We find that in some regime the theory should be described
in terms of a dual theory, very much in the spirit of
Seiberg duality in four dimensions. This duality does not coincide
with mirror symmetry.
}

\maketitle

\section{Introduction}

The past few years have seen a tremendous advance in our understanding
of supersymmetric gauge theories in various dimensions. Recently
interest has focussed on $N=2$ supersymmetric field theories in three dimensions
\cite{berc2,3d,berc1}.
Theories with double the amount of supersymmetry ($N=4$) in three dimensions
have allready been extensively studied and were found to posses a certain
duality, called mirror symmetry \cite{mirror, hananywit}. It interchanges Coulomb and Higgs branch
of the two theories which are mirror to each other.

Also many results have been obtained in theories with the same amount of 
supersymmetry, but with one more space dimension, that is $N=1$ theories in 
four dimensions (\cite{seiberg, pousp}). These theories have a duality symmetry which is
usually refered to as Seiberg duality. Two different gauge theories flow
to the same interacting IR fixed point. It is an interesting question, which
of those dualities survive the transition to $d=3$, $N=2$ and how they
turn out to be related if they do. Some progress was achieved in
\cite{berc1,berc2,3d}
where it was shown that at least mirror symmetry has a counterpart
in $d=3$, $N=2$ theories. Mirrors were explicitely constructed for
abelian gauge groups. They were shown to have an interpretation as
the theory of vortices in \cite{3d}. In \cite{3d,berc2} also the
tools were developed for the study of the non-abelian theories.
We will mostly stick to the notation of \cite{3d}.

The question whether Seiberg duality survives dimensional reduction
has not been answered so far. In this letter we will argue that there
actually exists a corresponding duality in the three dimensional
context. It basically leaves the Higgs/Coulomb structure untouched and
hence does not coincide with mirror symmetry. We will explicitly construct
this duality for $Sp(2N_c)$ \footnote{We will denote by $Sp(2N_c)$ the group
whose fundamental representation has dimensoin $2N_c$}
with $2N_f$ fundamental flavors,
since this seems to be the easiest case to
study. It can probably be generalized to other gauge groups as well.
In section 2 we will describe the model, analyze its behaviour
for small values of $N_f$ following \cite{3d} and construct
the dual. In section 3 we will describe some consistency checks:
the dual model captures the structure of Coulomb and Higgs branch,
reacts consistently under perturbations, reduces to the corresponding
Seiberg duality constructed in \cite{pousp} upon
decompactification and satisfies the
mod 1 parity anomaly matching conditions of \cite{3d}.
In Section 4 we will describe the brane point of view and conclude.

\section{The model}

The model we study is very similar to $d=3$, $N=2$ SUSY QCD with
$SU(n)$ gauge group which was extensively studied in \cite{3d}.
We refer the reader to their paper for a discussion of $N=2$
supersymmetry in 3 dimensions and the tools to analyze
the corresponding quantum theories. We will be dealing with
a $Sp(2N_c)$ gauge group instead. This has the advantage that
the only gauge invariant operators we have to consider are
the mesons. In $Sp$ theories the baryons are just
products of mesons. Let us first introduce the building blocks of
our analysis: we have $2N_f$ quarks Q transforming as fundamentals of
the gauge group\footnote{Even though in three dimensons theories
with an odd number of flavors would be anomaly free, too, with
the addition of suitable Chern-Simons terms, we will only consider
theories with an even number of flavors, since we do not
want to consider nonvanishing Chern-Simons contributions.}.
The only gauge invariant operator we can build from
them are the flavor antisymmetric mesons:
$$ M_{ij}=\omega_{ab}Q^a_i Q^b_j$$
where $a,b$ are color indices, $i,j$ are flavor indices and $\omega$ is
the 2-index antisymmetric invariant tensor of $Sp$. $M$ parametrizes
the Higgs branch of our model.

In addition we will have to deal with a scalar $\phi$
transforming in the adjoint representation of the gauge group.
It is part of the vector multiplet. If one
thinks of $N=2$, $d=3$ as dimensional reduction of $N=1$ in $d=4$ $\phi$
is the component of the 4d vector in the reduced direction. If
we are on the Coulomb branch, there is another scalar $\gamma$, which
is the dual of the vector. They can be combined into a complex scalar
$\Phi=\phi+i \gamma$. Let's write adjoint scalar in the form
$\phi= diag(\phi_1, \ldots, \phi_{N_c}, - \phi_1, \ldots, -\phi_{N_c})$.
By using the Weyl group we can take $\phi_1 \geq \phi_2 \geq \ldots \geq \phi_{N_c} \geq 0$.
Define semi-classically the fundamental instanton factors as
\begin{eqnarray*}
Y_j &\sim& e^{(\Phi_j-\Phi_{j+1})}, \; \;  j=1,\ldots, N_c-1 \\
Y_{N_c} &\sim& e^{2(\Phi_{N_c})}\\
\end{eqnarray*}
Again we expect that for $N_f \neq 0$ (like in \cite{3d}) the Coulmob branch is lifted
by instanton effects except for a one dimensional subspace. This is 
parametrized by an instanton factor $Y=\prod_{j=1}^{N_c} Y_j$ which is globally
well defined. To see how $Y$ transforms
under the global symmetries one has to count fermionic zero modes.
Using the formulas presented in \cite{berc2} it is easy to verify that
$Y$ picks up a $N_c$ factors of r-charge $-2$ from the gaugino zero modes in 
the $N_c$  fundamental instantons. 
In addition $Y_{N_c}$ and hence also $Y$ picks up one fermionic zero mode
from every massless quark present.
To summarize this, the following table shows
these building blocks and how they transform under the global symmetries.
Note that there are no anomalies for the global symmetries
in 3 dimensions, so we have the full classical symmetry.

\begin{displaymath}
\begin{array}{c| c c c}
     & U(1)_R & U(1)_A&  SU(2N_f)\\   
     \hline
	  Q & 0 & 1 &  {\bf 2N_f} \\ 
	  M & 0 & 2  & \tableau{1 1}\\ 
	  Y & 2(N_f-N_c) & -2N_f & {\bf 1} . \\
\end{array}
\end{displaymath}

The behaviour of the quantum symmetry depends crucially on the number of 
flavors:

\subsection{$N_f=0$:}
With no massless quarks present, all the fundamental instantons have
just r-charge $-2$. A superpotential
$$W=\sum_{j=1}^{N_c} \frac{1}{Y_j}$$
is created, where $Y_j$ denotes the jth fundamental instanton.
The full Coulomb branch is lifted and there is no stable supersymmetric
groundstate.

\subsection{$0 < N_f < N_c$:}
Since now one fundamental instanton picks up a quark zero mode 
contribution and hence can not appear in the 
superpotential, we expect a one dimensional sublocus of
the Coulomb branch (parametrized by $Y$) to remain unlifted. In terms of $Y$ and $M$
a superpotential is created, which is completely characterized by
the charge assignments (up to a normalization):
$$W= (N_c-N_f)(Y \Pf M)^{1/(N_f-N_c)}.$$
Again we are left without a stable vacuum.

\subsection{$N_f=N_c, N_c+1$:}
For $N_f=N_c$ a quantum constraint on the moduli space is
consistent with the global symmetries:
$$Y \Pf M =1.$$
The moduli space is a merged Higgs and Coulomb branch.

For $N_f=N_c+1$ the theory confines and the relevant degrees of freedom
are $Y$ and $M$ interacting via a superpotential:
$$W= -Y \Pf M.$$
In the 4d context this kind of behaviour was called s-confining (\cite{confine}).
All this is very similar to the phenomena observed for $SU(n)$ gauge group in
\cite{3d}. In their case the s-confining superpotential contained
a cubic term involving the baryons and hence led to a non-trivial fixed point.
The absence of the baryons makes the physics of the $Sp$ groups somewhat
easier to study.

Upon decompactification to 4 dimensions (which corresponds to
adding a term $\eta Y$ to the superpotential, with $\eta \sim 
e^{-1/Rg_3^2}$ (\cite{3d}) ) these results nicely connect to
the corresponding theories studied in \cite{pousp} (in the
4d limit $\eta \sim e^{-1/g_4^2} \sim \Lambda^{\beta_0}$,
$\beta_0=3 \mu_{gauge}- \mu_{matter}$ is the one loop beta function).
Note however that the sequence of the 3d theories is shifted
by one from the corresponding 4d theories (there one has:
no ground state for $N_f \leq N_c$, quantum constraint for
$N_f=N_c+1$ and s-confining for $N_f=N_c+2$).
For $N_c=1$ the picture described so far coincides with the $SU(2)$
analysis of \cite{3d}.

\subsection{$N_f \geq N_c+2$}
In the spirit of \cite{3d} we assume the following structure
emerging as we drive the theory towards higher values of
$N_f$. There will be a 1-dimensional Coulomb branch parametrized
by $Y$ intersecting the Higgs branch at $Y=0$. At the
intersection we get an interacting SCFT, best described by some
new degrees of
freedom.
We claim that in fact this setup has a dual description, very much
like the ones found in the 4 dimensional context. 
The dual theory has gauge group $Sp(2(N_f-N_c-1))$ and the following
content:

\begin{displaymath}
\begin{array}{c|c|c c c c}
     &Sp(2(N_f-N_c-1))& U(1)_R & U(1)_A&  SU(2N_f)& SU(2)\\
	  \hline
    q &\tableau{1}& 1 & -1 &  {\bf 2N_f}& \\
    t &\tableau{1}&N_c-N_f+1&N_f &{\bf 1} &2 \\
      M &1& 0 & 2  & \tableau{1 1}\\
	Y &1& 2(N_f-N_c) & -2N_f & {\bf 1} . \\
\end{array}
\end{displaymath}

and an aditional superpotential
$$W= Mqq+Ytt+\tilde{Y}.$$
$\tilde{Y}$ is the corresponding object to $Y$ of the magnetic theory. It picks
up global charges from gaugino zero modes of the dual group
and quark zero modes of q and t. The superpotential
removes the Coulomb branch associated with the magnetic gauge group. 
Note that the appearence of $\tilde{Y}$ in the superpotential
reduces the number of global $U(1)$s by one (as the other two terms
naturally do, too). This leaves us indeed with the same global symmetries
as we had in the original model. The fields $M$ and $Y$ appear as 
fundamentals in the dual theory, the magnetic invariants $qq$ and $tt$
get removed via the superpotential. Probably we should think of
the internal $SU(2)$ as being gauged, too, so that there are no invariants
$qt$. In the following we will present some evidence that this
dual picture indeed captures the physics of our original problem.

\section{Consistency Checks}
\subsection{The parity anomaly matchings}
As pointed out in \cite{3d} there is something similar to the 't Hooft 
anomaly matching conditions in three dimensions: whether or
not a weakly gauged global symmetry would require a Chern-Simons
term and hence break parity should match in both theories. Since this
is only a yes or no question (which amounts on checking if the
one-loop contribution is integer or half-integer), this check
is much weaker than the 4d equivalent, but nevertheless provides
us with some method to check our claims. In both theories we find that the 
corresponding Chern-Simons terms have coefficients (mod 1):
$k_{RR} =\frac{N_c}{2}$, $k_{AA}=0$, $k_{SU}=0$

\subsection{Classical Constraints on Higgs and Mixed branch}
Classically $M$ is constrained to have rank $\leq 2N_c$. On the magnetic
side $M$ appears as a singlet field and is classically unconstrained. 
For nonzero $Y$ in general we can still have nonvanishing $M$ giving
rise to mixed branches. In this case $rank(M) \leq 2(N_c-1)$, which
amounts to the classical condition of having at least one unbroken 
$U(1)$. It is very convincing to see how these constraints arise through
quantum effects in the dual description.

First consider the case $Y \neq 0$. The superpotential gives mass to $t$
and we are left with an $Sp(2(N_f-N_c-1))$ gauge theory with
the $2N_f$ fundamentals $q$ and the superpotential $W=Mqq+\tilde{Y}$.
Giving a vev to $M$ gives a mass to $\frac{rank(M)}{2}$ of the $q$.
If $rank(M)$ is to high the number of massless flavors drops into
the regime where we generate a superpotential and remove all supersymmetric
ground states. This puts a quantum constraint on the rank of $M$.
So far this discussion is identical to the $d=4$ case. In what follows
we will see that the $\tilde{Y}$ term plays a crucial role and
the consistency of the construction strongly supports the $d=3$ picture.
The $Mqq$ term in the superpotential tells us that the only possible
ground state is at the origin, $q=0$. In our previous analysis
we found that if $N_f^{L}$ (number of massless quarks, $N_f-\frac{rank(M)}{2}$)
is less or equal than $\tilde{N_c} (=N_f-N_c-1)$, the origin is removed from
the quantum moduli space. For $N_f^L  = \tilde{N_c}+1$ we
found that the quantum moduli space is the same as the classical moduli space
described in terms of $\tilde{Y}$ and $\Pf q^2$ interacting via
a superpotential:
$$W=\tilde{Y} \Pf q^2.$$
and hence includes the origin. With the additional superpotential term
$\tilde{Y}$ this becomes
$$W=\tilde{Y}(\Pf q^2 +1)$$
which again gives a smoothed out moduli space and removes the origin.
Hence we arrive at the following quantum constraint:
\begin{eqnarray*}
N_f^L &\geq& \tilde{N_c}+2 \\
rank(M) &\leq& 2(N_c-1) \\
\end{eqnarray*}
exactly reproducing the classical constraint of the
original theory.

For $Y=0$ we have an additional massless flavor, $t$, and hence
the above analysis gets modified to yield
$ rank(M) \leq 2N_c$, which again agrees with the classical constraint
obtained before.
\subsection{Decompactification to 4 dimensions}
In \cite{3d,seibwit} it was argued that decompatification to
4 dimensions can be achieved by taking into acount $d=3$ instantons
which are related by a large gauge transformation around the compact circle
to the usual instantons. Their effect is included by adding to the 
superpotential a term $\eta Y$ where $\eta \sim e^{-1/Rg_3^2}$
which becomes $0$ in the 3d limit and $\Lambda^{\beta_0}$ in
the 4d limit.
Adding this terms to the dual theory the equations of motion
for $Y$ tell us that:
$$<tt>=-\eta.$$
This breaks the internal $SU(2)$  and higgses the dual gauge group
to $Sp(2(N_f-N_c-2))$, which is precisely the expected
magnetic gauge group found in 4 dimensions (\cite{pousp}). The
$t$ fields get eaten by the Higgs mechanism and the $\tilde{Y}$
term in the superpotential gets removed by considering the effects
of decompactifying the magnetic gauge dynamics.
Hence we are only left with an 4 dimensional $Sp(2(N_f-N_c-2))$
gauge theory with $2N_f$ quarks $q$ and gauge singlets $M$
interacting via a superpotential.
$$W=Mqq.$$
The scale of the magnetic theory after the higgsing is
$\tilde{\Lambda} \sim \frac{1}{<tt>} \sim \frac{1}{\eta} \sim \frac{1}{\Lambda^{\beta_0}}$.
Our model reduces exactly to the well known Seiberg duality in
4 dimensions.

\subsection{Reduction to known results}
The model should be consistent under perturbations. While
adding real quark masses yields a rather complicated structure,
the effect of complex masses is easily studied.
In fact it is exactly the same as in 4 dimensions, since $M$ only
interacts via the $Mqq$ term in the superpotential.
Similarly perturbations along the $M$ flat directions can 
be analyzed. Note that other than in 4 dimensions
the cubic superpotential alone guarantees that we get
a non-trivial fixed point. We don't get a mapping to
free magnetic theories.

To get a new consistency check it is instructive to
study the breaking of $N_f=N_c+2$ to $N_f=N_c+1$. The
dual group is completely broken. Instanton contributions 
should reproduce the result presented in section 2.
Consider adding to the $N_f=N_c+2$ theory
a superpotential $W_{tree}=m M_{N_f+2,N_f+2}$. Assume
that the light mesons $\hat{M}$ are all nonzero. This
leaves the dual $Sp(2)$ theory with two quarks $q$ and $t$,
where $Sp(2)$ is then broken by the non-zero $<qq>=-m$.
At this stage the superpotential reads
$$W=\tilde{Y} tt <qq> +Y_L tt + \tilde{Y}$$
yielding a vev of $<tt>=-\frac{1}{<qq>}=\frac{1}{2m}$ (breaking
the internal $SU(2)$).
Since $Y_L$ can be associated with the scale of the low energy
theory (after integrating out the quarks that get masses $<\hat{M}>$)
we expect $Y_L \sim Y \Pf<M>$. Finally we reproduce as expected
the $N_f=N_c+1$ superpotential
$$W=Y \Pf \hat{M} $$ 

\section{Brane Interpretation and Conclusions}

In the brane setup this theory can be
obtained by rotating a NS5 brane in the Hanany-Witten setup 
\cite{hananywit} or by T-dualizing the EGK \cite{EGK} setup.
The first approach corresponds to breaking $N=4$ to 
$N=2$ and the second one to reducing $N=1$, $d=4$ down to
three dimensions. Viewing things this way makes
mirror symmetry a natural consequence of the S-duality of Type IIB.
On the other hand Seiberg duality was obtained in a
rather complicated fashion by pushing around the branes of EGK.
As discussed in \cite{vafa} it is not even clear if this really
derives Seiberg duality from brane dynamics. It
is not possible to decide from this description, whether Seiberg survives
reduction to 3d or not. But at least we would not expect
that Seiberg duality should reduce to mirror symmetry,
since latter is allready taken care of by S-duality. This
is consistent with what we presented in this paper.

The brane picture may also explain, where the shift
in the dual gauge group comes from (note that instead
of $Sp(2(N_f-N_c-2))$ in 4d we have $Sp(2(N_f-N_c-1))$ in 
3d). In 4d the $-2$ was obtained in the various approaches
\cite{ zwieb, vafa} by pushing branes through orientifold planes.
Charge conservation required the destruction of two branes in
certain transitions to cancel the charge of the orientifold,
giving rise to the above $-2$ factor. Going to 3 dimensions
via T-duality reduces the orientifold by one dimension and
hence halves it charge (\cite{polchinski}), substituting
$-2$ with $-1$.

We expect similar dualities to be valid for other gauge groups, too.
Their structure is certainly more complicated due to the presence of baryons.
We hope to be able to report on progress in those direction soon.

\section*{Acknowledgements}

We are very greatful to D. L\"ust and I. Brunner for useful
discussions on the subject.

\end{document}